\documentclass[manuscript]{aastex}

\slugcomment{Version of Sept 24 2008}

\shorttitle{New White Dwarfs in UKIDSS}
\shortauthors{Lodieu et al.}

\begin{document}

\title{Cool White Dwarfs Identified in the Second Data Release of the 
UKIRT Infrared Deep Sky Survey}


\author{N. Lodieu\altaffilmark{1}}
\author{S. K. Leggett\altaffilmark{2}}
\author{P. Bergeron\altaffilmark{3}}
\and
\author{A. Nitta\altaffilmark{2}}

\altaffiltext{1}{Instituto de Astrof\'\i sica de Canarias, C/ V\'\i a L\'actea s/n,
E-38205 La Laguna, Tenerife, Spain }
\altaffiltext{2}{Gemini Observatory, Northern Operations Center, 670
  N. A'ohoku Place, Hilo, HI 96720, USA}
\altaffiltext{3}{D\'epartement de Physique, Universit\'e de Montr\'eal,
C.P.\ 6128 Succursale Centre-Ville, Montr\'eal, QC H3C 3J7, Canada}

\begin{abstract}
We have paired the Second Data Release of the Large Area Survey of the
UKIRT Infrared Deep Sky Survey with the Fifth Data Release of the
Sloan Digital Sky Survey to identify ten cool white dwarf candidates,
from their photometry and astrometry.  Of these ten, one was
previously known to be a very cool white dwarf. We have obtained optical
spectroscopy for seven of the candidates using the GMOS-N spectrograph
on Gemini North, and have confirmed all seven as white dwarfs. Our
photometry and astrometry indicates that the remaining two objects are
also white dwarfs. Model analysis of the photometry and available
spectroscopy shows that the seven confirmed new white dwarfs, and the
two new likely white dwarfs, have effective temperatures in the range $T_{\rm
eff}=5400-6600$~K. Our analysis of the previously
known white dwarf confirms that it is cool, with $T_{\rm eff}$ =
3800~K.  The cooling age for this dwarf is 8.7 Gyr, while that of the 
nine $\sim$ 6000~K white dwarfs is 1.8--3.6 Gyr. We are unable to determine
the masses of the white dwarfs from the existing data, and therefore 
we cannot constrain the total ages of the white dwarfs.  The large cooling age 
for the coolest white dwarf in the sample, combined with its low
estimated tangential velocity, suggests that it is an old member of  the
thin disk, or a member of the thick disk of the Galaxy, with an age 10--11~Gyr. 
The warmer white dwarfs appear to have velocities typical of the thick disk 
or even halo; these may be very old remnants of low-mass stars, or they
may be relatively young thin disk objects with unusually high space motion.   
\end{abstract}



%
%%%%%%%%%%%%%%%%%%%%%%%%%%%
%%%% Introduction %%%%
%%%%%%%%%%%%%%%%%%%%%%%%%%%
%
\section{Introduction}

White dwarfs are the end stage of stellar evolution for the vast majority 
of stars --- all stars less massive than 8 $M_{\odot}$ end their lives as 
cooling white dwarfs. The coolest white dwarfs can therefore constrain the 
age of the Galactic disk, or even of the halo, if such objects can be found. 
Most white dwarfs consist of a C/O core with an outer envelope composed of 
helium and/or hydrogen, with occasional traces of metals. 
The ratio of the number of hydrogen-rich to helium-rich white dwarfs
is a function of  $T_{\rm eff}$, and the
chemical evolution of white dwarf atmospheres is complex
(e.g.\ Bergeron, Leggett \& Ruiz 2001; Tremblay \& Bergeron 2008).  
The mass and composition of both the core and the atmosphere controls the 
cooling rate of the white dwarf. Bergeron et al. (2001) use atmospheric and 
evolutionary models to analyse a sample of white dwarfs with measured 
trigonometric parallaxes to show that the coolest of these 
white dwarfs, with $T_{\rm eff} \sim$4000--4500~K, are 9--10 Gyr old 
if they have a thick hydrogen atmosphere, and 8--9 Gyr old if they have a 
helium-rich atmosphere. These ages are consistent with the age of the 
local Galactic disk (e.g.\ Leggett et al.\ 1998).

Several groups are trying to find even cooler and older white dwarfs in 
order to confirm the age of the disk, and to investigate the ages of older 
Galactic components. The {\it Hubble Space Telescope} has enabled the 
detection of white dwarf cooling sequences in clusters; 
Hansen et al. \ (2007) has recently identified hydrogen-rich white dwarfs 
with cooling ages of 11~Gyr at the truncation of the white dwarf sequence 
in the 11.5~Gyr old globular cluster NGC 6397. Oppenheimer et al.\ (2001) 
identified a sample of 
high-velocity white dwarfs which was inferred to be a halo population by 
their kinematics. However Reid et al. \ (2001) suggest that the majority 
of this sample has kinematics consistent with thick disk membership, and
analysis of the sample by Bergeron et al.\ (2005) 
found that the white dwarfs were relatively warm, implying relatively short 
cooling ages. The age of the Oppenheimer et al. sample remains a matter of 
debate (e.g.\ Ducourant et al.\ 2007 and references therein).

Very cool white dwarfs are unambiguously old,
as their total age is dominated by the large cooling time. Such
white dwarfs have been found in the Sloan Digital Sky Survey 
(SDSS, York et al.\ 2000). Kilic et al.\ (2006) use the SDSS and 
US Naval Observatory catalog (USNO-B; Monet et al.\ 2003) to identify 
cool white dwarfs using 
a reduced proper motion (RPM) diagram. The RPM is defined as 
\begin{equation}
\label{eq_flux}
H_{\rm mag} = \rm mag + (5 \times {\rm log(\mu)}) + 5
\end{equation}
where the proper motion $\mu$
is measured in arcseconds per year. Here the apparent magnitude (mag) and
$\mu$ are used as a proxy for absolute magnitude for a sample with similar 
kinematics (see e.g. Jones 1972). 

Kilic et al. (2006) have spectroscopically confirmed several white dwarfs 
with $ 15.7 < r < 19.7 $
using the RPM diagram, including sixteen with $T_{\rm eff}$ around 
4000~K or cooler. The RPM diagram was also used by 
Carollo et al.\ (2006) to identify cool white dwarf candidates in the Guide 
Star Catalog II  (GCS-II) database, of which 24, with $ 15 < R_{\rm F} < 20 $, were confirmed by spectroscopy 
to be previously unknown white dwarfs.  Hall et al.\ (2008) recently 
identified an $r = 18.8$  halo white dwarf candidate in the SDSS from its spectrum and 
high proper motion. Vidrih et al.\ (2007) also used the RPM diagram 
to identify over 1000 cool white dwarf candidates in a deeply 
imaged SDSS region known as Stripe 82, including 24 candidates that may 
be cooler than 4000~K, and 34 halo white dwarf candidates. These candidates,
which have  $ 18 < r < 22 $,
are yet to be confirmed spectroscopically.

Low-temperature hydrogen-rich white dwarf atmospheres are at high pressures, 
and show strong pressure-induced molecular hydrogen (H$_2$) opacity. This 
opacity has a broad absorption feature around 2 $\mu$m, which affects the 
$H$ and $K$ near-infrared bands, centered near 1.65 and 2.2 $\mu$m, 
respectively. When $T_{\rm eff}$ decreases below 4000~K, the opacity also 
impacts the red (0.8 $\mu$m) and far-red (1.1 $\mu$m) colors (Borysow 2002). 
Harris et al.\ (2001) and Gates et al.\ (2004) used this feature 
to identify and confirm six extremely cool white dwarfs, with
$ 18.9 \leq r \leq 19.6 $,
 by their unusual 
SDSS colors. Harris et al.\ (2008) extend this study and present an 
additional seven cool white dwarfs with $ 18.7 \leq r \leq 20.4 $
found in the SDSS by their 
colors. Rowell et al.\ (2008) in a similar way identified an 
$ R_{\rm F} = 17.8 $ ultracool 
white dwarf in the SuperCOSMOS Sky Survey (Hambly et al.\ 2001)
from its $BRI$ colors.

Although analysis of white dwarfs with such cold and high-pressure 
atmospheres is 
difficult (e.g.\ Bergeron \& Leggett 2002) it is important to add to the 
still small sample of cool and old white dwarfs.  Not only do these objects 
impart information about the history of the Galaxy, but improving our 
understanding of the physics of such atmospheres is of general significance, 
for example for modelling cool high-pressure planetary and brown dwarf 
atmospheres.

In this paper we present the results of a search of the UKIRT Infrared Deep 
Sky Survey (UKIDSS; Lawrence et al.\ 2007) for cool white dwarfs. We 
identified our sample by pairing the optical photometry given in the SDSS Data Release Number 5 
(DR5; Adelman-McCarthy et al.\ 2007) with the infrared photometry in Data Release Number 2 of the Large 
Area Survey (LAS) of UKIDSS (DR2; Warren et al.\ 2007b). Both techniques 
described above were utilized --- candidates were selected using the 
RPM diagram, and also by color, selecting for the presence of H$_2$ 
opacity.  
The following sections describe the LAS (\S 2), the sample 
selection (\S 3), the results of our spectroscopic follow-up (\S 4)
and model analysis (\S 5), and a discussion of these results (\S 6).  
We show that we have discovered faint and cool white dwarfs
with  $ 19.5 \leq r \leq 20.6 $ and $ 5400 \leq T_{\rm eff}$~(K)~$\leq 6600$.

%
%%%%%%%%%%%%%%%%%%%%%%%%%%%
%%%% UKIDSS %%%%
%%%%%%%%%%%%%%%%%%%%%%%%%%%
%

\section{The UKIDSS Large Area Survey}

UKIDSS (Lawrence et al. 2007) is a large-scale infrared survey conducted
with the UK InfraRed Telescope (UKIRT) Wide Field Camera
(WFCAM; Casali et al.\ 2007). WFCAM uses filters following the Mauna Kea
Observatories specification (Tokunaga, Simons \& Vacca 2002), and
the UKIDSS photometric system is described by Hewett et al.\ (2006).
Observations began in May 2005. The data and catalogues generated
by the automatic pipeline processing can
be retrieved through the WFCAM Science Archive (WSA; Hambly et al.\ 2008).

All data are pipeline-processed by the Cambridge Astronomical Survey Unit
(CASU; Irwin et al., in preparation) following a standard procedure for
infrared images. An extensive description of each step involved in the 
processing of the WFCAM data is available on the CASU 
webpage\footnote{http://casu.ast.cam.ac.uk/surveys-projects/wfcam/technical}.
Summaries of the data reduction can also be found in Lawrence et al.\ (2007),
Dye et al.\ (2006), and Warren et al.\ (2007a).

UKIDSS actually consists of five survey components, one of which is the
Large Area Survey (LAS). The LAS is the sub-survey most likely to contain 
faint and rare sources of the local Galaxy, such as the cool white dwarfs 
and brown dwarfs.  The LAS aims to survey 4000 square degrees in $YJHK$
with a second epoch at $J$, to reach $J \sim$ 20 mag. The $5 \sigma$ photometric
depths of the second Data Release of the LAS are $Y = 20.2$, $J = 19.6$,
$H = 18.8$, and $K = 18.2$ mag (Warren et al.\ 2007b). The area surveyed 
by the LAS was designed to overlap with the SDSS, divided up into three
blocks. The equatorial block with  Right Ascension 23 to 04 hours 
and Declination between $-$1.5 and $+$1.5 degrees overlaps SDSS 
stripes 9 to 16. The southern block covers 8 to 14 hours 
and  (approximately) $-$3 to $+$15 degrees, and includes  
SDSS stripe 82. Finally, the northern block  
(available in upcoming data releases) will provide an overlap 
with SDSS stripes 26 to 33. This information is 
detailed in Lawrence et al.\ (2007), Dye et al.\ (2006), and Warren
et al.\ (2007a).

A significant amount of multi-band photometric data has already
been released worldwide: the Early Data Release (EDR; Dye et al.\ 2006)
and Data Release Number 1 (DR1; Warren et al.\ 2007a). In addition,
a second Data Release was made available to the ESO community in March
2007 (DR2; Warren et al.\ 2007b), a third  in December 
2007 (DR3), and a fourth in July 2008. The sources presented 
here were selected in July 2007 from LAS DR2, which included 
282 deg$^2$ of $YJHK$ data.

%
%%%%%%%%%%%%%%%%%%%%%%%%%%%
%%%% Sample selection %%%%
%%%%%%%%%%%%%%%%%%%%%%%%%%%
%
\section{Sample Selection}

We have used Structured Query Language (SQL) and the  WFCAM Science Archive
to carry out a cross-correlation of the UKIDSS LAS DR2 and SDSS DR5 databases.
We have restricted our queries of the LAS database to detections classified
as point sources 
(``mergedClass'' parameter equal to $-1$
\footnote{A classification code where a point source has a 
value of $-$1 and a galaxy $+$1. A full description is available at
http://surveys.roe.ac.uk/wsa/www/gloss\_m.html\#lassource\_mergedclass.})
and to good detections only ({``ppErrBits'' $<$256
\footnote{A value describing the quality of the detection.
More details on this parameter are available at 
http://surveys.roe.ac.uk/wsa/www/gloss\_y.html\#lassource\_ypperrbits.})
to avoid cross talk\footnote{Cross-talk artefacts are due to the
presence of a nearby bright star.} and other artefacts. 
Similar queries were used to find brown dwarfs in the field and in open
clusters (Lodieu et al.\ 2007a, b, c); additional examples and details can
be found in Hambly et al.\ (2008). The query imposed a detection in all of
$YJH$, and included color cuts as well as a lower limit on the proper motion,
as described below. Sources were matched by requiring the presence of a
``primary'' SDSS source within 2$\arcsec$ of the LAS coordinates. 
Increasing the search radius to 5$\arcsec$  picked up either no additional source, or additional sources that were clearly matched to other UKIDSS sources.
The query returned coordinates, 
photometry and errors from both surveys, as well as the proper motion 
(Tables 1 and 2).

The proper motion was computed from the difference in the LAS DR2 and
SDSS DR5 coordinates. The WFCAM astrometry is  tied to the
2MASS point source catalogue and has a systematic accuracy of $< 0\farcs 1 
\ rms$ (Dye et al. 2006).  Figure 1 shows the size of the scatter in the
difference between the SDSS DR5 and LAS DR2 astrometry, as a function of $J$-band brightness.
For the sources considered here, with $ 18.7 \leq J \leq 19.5$, the typical 
uncertainty is $0\farcs 025$ to $0\farcs 04$. 
The  LAS and SDSS epochs differ by 2 to 7 years for all our candidates, 
and 2 to 4 years for our confirmed white dwarfs. Hence our lower limit to proper motion of
0.1 arcsec yr$^{-1}$ can be measured to $>5 \sigma$.
We note that the proper motion measured for the brightest white dwarf in our sample using  the SDSS DR2 and USNO-B catalogs
is in good agreement with our value --- the WFCAM data gives ($\mu_\alpha$, $\mu_\delta$) of ($+$0.142, $-$0.084) cf. 
($+$0.128, $-$0.072) arcsec yr$^{-1}$ (Kilic et al. 2006).

Our color selections were based on a combination of published and modelled
colors of cool ($T_{\rm eff} < 7000$~K) hydrogen-rich white dwarfs, as available in July 2007. An initial search using Bergeron et al. (1995) model colors only, for   $T_{\rm eff} \leq 4000$~K produced no matches.
We applied the following selections to the SDSS 
DR5 and LAS DR2 databases, where the $gri$ are AB magnitudes and the
$JHK$ are Vega magnitudes:
$$ +0.2 \leq g - r \leq +1.2 $$
$$ -0.6 \leq r - i \leq +0.6 $$
$$ J - H < -0.1 $$
$$ H - K < -0.1 $$
$K$ non-detections (i.e. $K > 18.2$) were also included if the $g - r$, $r - i$ and $J - H$ criteria were met.
To avoid saturation we  selected objects fainter than $J = 14.0$;
the lower limit on $J$  is set by the requirements of a blue $J - H$ color
and a detection at $H$.  We selected objects fainter than the 
$5 \sigma$\ $H$ detection limit of 18.8; we identified objects as faint as 
$H=19.7$ although that results in highly uncertain $H$ magnitudes, as we
discuss further below. 
 
This query was designed to pick up neutral to red stellar sources in $g-r$, 
with blue near-infrared colors indicating pressure-induced H$_2$ opacity in the near-infrared. Very red sources ($ g-i > 1.8$) were excluded as these are likely to be subdwarfs for the reduced proper motion values considered here
($H_{\rm g}>20$, see e.g. Figure 1 of Kilic et al. 2006). Thus our search is designed to find cool sources with high-pressure hydrogen-rich atmospheres.
The query returned 586 objects.

The sample size was reduced by requiring a proper motion larger than $0\farcs 1$ yr$^{-1}$ (corresponding to a $> \sim$5$\sigma$ detection on the total  
motion, as described above). Our sample 
has  $g = 20 - 21$, hence this proper motion selection implies
reduced proper motions $H_g > 20$, appropriate for discovering previously 
unrecognized white dwarfs in the old disk and halo (cf.\ Figure~1 of Kilic 
et al.\ 2006). The proper motion cut reduced the sample to ten objects. 
Of these ten objects, seven were accessible over the allocated telescope 
time period and had no spectra, one had already been identified and 
observed by Kilic et al.\ (2006) and two others were inaccessible. 
Table 1 lists the astrometry and reduced proper motion for all ten 
candidates, and Table 2 gives their SDSS and LAS photometry from
the latest releases, i.e.\ SDSS DR6 and LAS DR3.

Our databases search did not pick up any of the other recently identified
cool white dwarfs described in \S 1, either because the sources
were outside the LAS DR2 sky area or the color criteria (usually the
blue near-infrared colors) were not met. For example, only three
out of the 112 white dwarfs reported by Kilic et al.\ (2006) are
detected in $YJH$ and lie within the LAS DR2 area, and only one of those
met our color criteria.
Other sources not in our sky area include those of Gates et al.\ (2004),
Hall et al.\ (2008), Rowell et al.\ (2008), and most of the
Carollo et al.\ (2006) and Harris et al.\ (2008)
sources. Those with inappropriate colors include the Harris et al.\ (2001)
source that is not detected at $J$ or $H$; two Carollo et al.\ sources;
and two sources from Harris et al.\ (2008), one of which is not detected at $H$ and the
other of which has $J - H >$ 0.  Of the 13 faint disk and halo candidates
identified by Vidrih et al.\ (2008) that are detected at all of $YJH$,
only one has $J - H < -$0.1, and that is the Kilic et al.\ white dwarf 
recovered in our search.

Figure 2 shows $g - r$:$r - i$ and $i - J$:$J - H$ color-color plots
demonstrating the location of all 586 preliminary candidates, as well the
ten final white dwarf candidates, and a main sequence drawn
from a sample of SDSS$+$LAS sources with small photometric errors.
Also shown are modelled colors from Holberg \& Bergeron (2006;
see also \S 5.1 below), and some of the recently published
cool white dwarfs (or white dwarf candidates) described above. 
Our color selections are indicated. While the $gri$ selection picks up both
warm to cool stars and white dwarfs, the cut $J - H < -0.1$ should eliminate 
all but the hydrogen-rich white dwarfs with  $T_{\rm eff} < 4000$~K,
according to the models.  An important caveat is that selecting candidates
that are faint and blue in the near-infrared produces
sources that are very faint at $H$, and
so the errors in $J - H$ are significant (see Figure 2 and Table 2).
The uncertainties in the fainter half of the sample with $H > 19.0$ may
also be underestimated; extrapolation of the uncertainties for the brighter sources suggests that these should have $\sigma H \sim 0.5$~mag cf. 0.3 mag.  We address this further in the data analysis presented in \S 5.

Figure 3 shows the $H_{\rm g}$:$g - i$ RPM diagram for our sample, and other
white dwarfs taken from the literature, as described in the caption.
The expected locations of the white dwarf cooling curves for the disk and
halo are indicated on the plot. Our confirmed white dwarfs lie at
the lower end of the white dwarf sequence and have $H_{\rm g} > 20.35$. 
The objects selected by our color and proper motion cuts do not form a
complete sample in  $H_{\rm g}$ space.  Faint objects with proper motion 
less than $0\farcs 1$  yr$^{-1}$ also lie in the region defined by
$H_{\rm g} > 20.35$; however targets with $r > 20.5$ were impractical for
followup in the allocated telescope time, and some targets were unreachable. Our sampling of the 586 SDSS$+$LAS targets with $H_{\rm g} > 20.35$ is 90\% complete for the sources with
$r < 20.3$ but only 25\% complete for those with $ 20.3 < r < 20.5$.

%
%%%%%%%%%%%%%%%%%%%%%%%%%%%
%%%% Spectroscopy %%%%
%%%%%%%%%%%%%%%%%%%%%%%%%%%
%
\section{Spectroscopic Observations}

Ten hours of Gemini North observing time was granted to this project through 
program GN-2007B-DD-6. Table 3 gives the total on-source exposure time for 
the seven targets, and the dates on which they were observed. We 
obtained long-slit spectroscopy with the GMOS-N instrument 
(Hook et al.\ 2004) during dark photometric and non-photometric conditions. 
The $1\arcsec$ slit was used with the 
R150 grating providing a resolution of $\sim 17$~nm, for the typical 
delivered seeing of $0 \farcs 7$ FWHM.  
We  blocked second-order contamination from 
wavelengths shorter than $\sim$450 nm by using the G0305 filter. The 
wavelength coverage obtained was 460 -- 950 nm, however detector fringing 
affected the spectra longwards of 820 nm. For this initial investigation of
the SDSS$+$LAS candidate list we chose wide wavelength coverage, and so low
resolution, with good coverage of the red.  The wide wavelength coverage
ensured that we could confidently identify subdwarfs or other 
non-white dwarf contaminants, and the extension to the red 
allowed detection of far-red pressure-induced H$_2$ effects, should any be present.  

Flatfielding and wavelength 
calibration were achieved using lamps in the on-telescope calibration unit. 
The standard star HZ 44 was used to determine the instrument response curve, 
and flux calibrate the spectra. The data were reduced using routines 
supplied in the IRAF Gemini package.

Figure 4 shows the GMOS spectra obtained by us, as well as the spectrum 
obtained at the Hobby-Eberly Telescope by Kilic et al.\ (2006) for 
SDSS J2242$+$00. For reference, spectra of an F dwarf and F subdwarf 
are also shown (taken from the spectral atlas of Le Borgne et al. 2003). 
Six of our seven objects show hydrogen lines pressure-broadened by the high 
gravities typical of white dwarfs, and no other features, while the 
seventh is featureless. Hence all seven of our observed candidates 
are confirmed to be white dwarfs.  

%
%%%%%%%%%%%%%%%%%%%%%%%%%%%
%%%% Modelling %%%%
%%%%%%%%%%%%%%%%%%%%%%%%%%%
%
\section{Modelling the Observed Colors and Spectra}
\subsection{Description of the Models and Fitting Technique}

The model atmospheres used in this analysis are described at length in
Bergeron et al.~(1995, with updates given in Bergeron et al.~2001,
2005). These models are in local thermodynamic equilibrium, they allow
energy transport by convection, and they can be calculated with
arbitrary amounts of hydrogen and helium.  Synthetic colors are
obtained using the procedure outlined in Holberg \& Bergeron (2006)
based on the  Vega fluxes taken from Bohlin \& Gilliland (2004).

The method used to fit the photometric data is similar to that
described in Bergeron et al.~(2001), which we briefly summarize here. We
first transform the magnitudes at each bandpass into observed average
fluxes $f_{\lambda}^m$ using the following equation

\begin{equation}
\label{eq_flux_obs}
m= -2.5\log f_{\lambda}^m + c_m
\end{equation}

\noindent
where the values of the constants $c_m$ for the infrared $YJHK$
photometry are obtained using the transmission functions from Hewett 
et al.\ (2006) and the Vega fluxes discussed above; we obtain $c_Y=-23.10069$,
$c_J=-23.81578$, $c_H=,-24.84612$, and $c_K=-26.00940$.  For the
optical $ugriz$ photometry, we simply rely on the definition of the
AB$_\nu$ magnitude system (see, e.g., eq.~3 of Holberg \& Bergeron
2006). Small corrections to the SDSS $uiz$ (not to $gr$) magnitudes
have been applied and included in the modelling following the 
work by Eisenstein et al.\ (2006).
The resulting energy distributions are then fitted with the
model Eddington fluxes $H_{\lambda}^m$ properly averaged over the
appropriate filter bandpasses (for the $ugriz$ system, we use the
transmission functions discussed in Holberg \& Bergeron 2006 and
references therein). The average observed and model fluxes are related
by the equation

\begin{equation}
\label{eq_flux_model}
f_{\lambda}^m= 4\pi~(R/D)^2~H_{\lambda}^m
\end{equation}

\medskip
\noindent where $R/D$ is the ratio of the radius of the star to its distance 
from Earth. Our fitting procedure relies on the nonlinear
least-squares method of Levenberg-Marquardt, which is based on a
steepest descent method. The value of $\chi ^2$ is taken as the sum
over all bandpasses of the difference between both sides of equation
(3), properly weighted by the corresponding observational
uncertainties. 
Since our models do not include the red wing opacity from Ly~$\alpha$
calculated by Kowalski \& Saumon (2006), we 
neglect here the $u$ bandpass in our fitting procedure since
this opacity may be important in the ultraviolet region.
We also neglect the $H$-band data due to the large uncertainties in 
these data. For one of our white dwarfs, ULAS J1522$+$08, the (faint) 
$z$-band magnitude appeared discrepant, compared to both other wavelengths 
and to the models, and was ignored. 

We consider only $T_{\rm eff}$ and the solid angle $\pi(R/D)^2$ free
parameters. The uncertainties of $T_{\rm eff}$ and the solid angle are
obtained directly from the covariance matrix of the fit. Since the
distance to each object in our sample is not known, we assume a
value of $\log g = 8.0$ in the following analysis.  White dwarfs have been
shown to have a very strongly peaked mass and surface gravity
distribution (e.g.\ Bergeron et al.\ 1992; Liebert, Bergeron \& Holberg 2005;
Kepler et al. 2008). DA white dwarfs have a mean mass of 
$0.6 \pm 0.1\ M_{\odot}$ while DBs are slightly more massive with
$0.7 \pm 0.1\ M_{\odot}$; these ranges infer a likely range in gravity
for our sample of $7.7 \leq \log g \leq 8.3$.

Figures 5 through 7 show the model fits to the observational data, assuming
 $\log g = 8.0$.  For most of the sample, relatively warm temperatures are 
derived of $T_{\rm eff} \approx$ 6000~K; for these the uncertainty in 
$T_{\rm eff}$ due to the photometric scatter is around 180~K 
(Figures~5 and 6). Experiments including the $H$-band data in the fits, 
both with the nominal photometric uncertainty, and with twice the nominal 
uncertainty (as might be expected for the faintest objects, based on an 
extrapolation of the S/N of the brighter objects), gave differences in 
derived temperature of only $\sim 30$~K; hence the uncertainty in
$T_{\rm eff}$ is dominated by the photometric scatter. 
(Fits to the spectral energy distributions with $\Delta T_{\rm eff} = 400$~K, i.e.
around twice the error derived from the scatter, produce synthetic fluxes that
fall well outside the error bars for the  $z$, $Y$ and $J$ datapoints.)
All the $H$-band datapoints 
appear faint because of our selection for objects with apparently blue $J - H$ color.
The faint magnitudes and associated large photometric errors have scattered 
relatively warm white dwarfs into our target selection. 

The previously known dwarf SDSS J2242$+$00 recovered in our selection, however, is a 
low-temperature white dwarf. This object is much cooler than the rest of the
sample, and we show below we can produce a good fit with $T_{\rm eff} = 3820\pm100$~K
(Figure~7). For this object both the $u$ and $g$ photometry 
was ignored, due to the missing Ly~$\alpha$ opacity which has a larger impact at
lower temperatures (e.g. Figure\ 4 of Kowalski \& Saumon 2006).

\subsection{Surface Gravity, Composition and  Temperature}

Since the energy distributions are not particularly sensitive to
$\log g$ in the temperature range considered
here, our assumption of $\log g = 8.0$ for all objects will not affect our
$T_{\rm eff}$ estimates. For instance, a variation of $\pm0.5$ dex in
$\log g$ yields differences in effective temperature of $\pm 15$~K on
average.  This is much smaller than the uncertainty due to the 
photometric variations.

The effect of the presence of helium on the predicted energy
distributions and spectra of DA stars in this temperature range
is discussed in detail in Bergeron et al.\ (1997; see their
\S 5.4 and Figures\ 23 and 24). 
Note that He~I lines become spectroscopically invisible 
for $T_{\rm eff} < 10000$~K, and so we would not detect helium features
in our sample.  While an atmospheric
composition of $N({\rm He})/N({\rm H})=1$ will not affect the energy
distribution and thus the temperature estimates significantly for 
$T_{\rm eff} \approx 6000$~K, the
H$\alpha$ line profiles are predicted to be much more shallow than in the 
pure hydrogen models. Hence the sharpness of the H$\alpha$ absorption
profiles reported here for six of the white dwarfs imply that these 
objects have hydrogen-rich atmospheres. The differences in 
$T_{\rm eff}$ that would be derived for the pure-helium fit range from 20~K 
to 120~K, for these white dwarfs with 5400 $< T_{\rm eff} <$ 6600~K 
(Figures~5 and 6). For the seventh white dwarf, ULAS J0302$+$00,
the lack of hydrogen features similarly constrains the atmosphere to be helium-rich. 
In this case however the difference in temperature between the two 
composition fits is only 20~K (Figure~5).

Neither pure-hydrogen or pure-helium atmospheres produced a good fit
to the energy distribution of the very cool white dwarf SDSS
J2242$+$00, discovered by Kilic et al.\ (2006). Instead, a good fit
was found using a model with almost identical amounts of hydrogen and
helium (Figure~7). In this case the featureless spectrum does not
constrain the composition, as hydrogen lines would not be present at
this low a temperature.

For the two white dwarf candidates without spectra, ULAS J1528$+$06 and 
ULAS J1554$+$08, Figure 6 shows that, 
if white dwarfs, these objects are relatively warm with 
6060 $\leq T_{\rm eff} \leq$ 6330~K. The faintness of the sources, 
combined with their significant proper motion, suggests that these objects 
are indeed evolved white dwarf remnants.

Table 4 lists the derived atmospheric properties of the ten white dwarfs 
discovered or recovered in our search of DR2 of the UKIDSS LAS. Using the 
composition and temperature, and assuming that these stars have the canonical 
white dwarf mass of 0.6 $M_{\odot}$, we can use the synthetic colors of 
Holberg \& Bergeron (2006, an extension of Bergeron, Wesemael \& Beauchamp 
1995) and the evolutionary sequences of Fontaine, Brassard \& Bergeron (2001) 
to derive both a cooling age and distance, and hence tangential velocity. 
These values are also given in Table 4, together with the uncertainty in 
$T_{\rm eff}$ -- due to photometric scatter -- as well as that in the implied 
cooling age, distance and velocity -- all of which are primarily due to the
uncertainty in gravity (or mass). We discuss the implications of 
these findings below.

%
%%%%%%%%%%%%%%%%%%%%%%%%%%%
%%%% Discussion %%%%
%%%%%%%%%%%%%%%%%%%%%%%%%%%
%
\section{Discussion}

Nine of the sample of ten white dwarfs found in our search have 
5400 $< T_{\rm eff} <$ 6600~K. The evolutionary models imply that their 
cooling ages are 1.8 -- 3.6 Gyr if they are 0.6 $M_{\odot}$ white dwarfs,
and around 2 Gyr older or 1 Gyr younger if they are more or less massive
(see \S 5.1 and Table 4).  The range in gravity used here of $\pm 0.3$~dex
corresponds to a range in mass of $\sim\pm 0.2 M_{\odot}$.
Recent studies of the initial-final mass relation
(Catal\'{a}n et al. 2008, Kalirai et al. 2008) suggest that low-mass stars can 
produce relatively high-mass white dwarf remnants.  Specifically, a
1.0 $M_{\odot}$ star will produce a 0.50  $M_{\odot}$ white dwarf, and
a 2.0 $M_{\odot}$ star will produce a  0.60  $M_{\odot}$ white dwarf.
As the main sequence lifetimes of such stars are 10 -- 2 Gyr, it is impossible 
to constrain the total age of our nine $\sim$6000~K white dwarfs; if they have
the canonical white dwarf mass the total age is  $\sim$5~Gyr, however if they
are even slightly less massive they may be much older.

The tenth source discovered by Kilic et al.\ (2006), SDSS J2242$+$00, is brighter, closer and cooler than the other objects in the sample. The evolutionary models give it a 
cooling age of 8.7 Gyr.  Hence SDSS J2242$+$00 is clearly old with a total age $>$9 Gyr.

The models and data allow us to estimate distances and tangential velocities
for the white dwarfs in our sample (Table 4).
Bergeron, Ruiz \& Leggett (1997, their Figure 34) and Holberg, Bergeron \& Gainninas (2008) show that distances determined using absolute model fluxes,
with model parameters determined either from spectroscopy or photometry,
agree well with those measured trigonometrically.
The LAS sample of white dwarfs can probe to fainter SDSS magnitudes 
and hence greater distances than for example the Kilic et al.\ (2006) sample. 
Kilic et al.\ required good detections in USNO-B in order to determine 
reliable proper motions, and hence was limited to $g <$ 20 mag 
(Monet et al.\ 2003). The LAS sample goes one magnitude fainter and 
includes sources with 19.7 $< g <$ 21.0 mag. The cool white dwarf SDSS J2242$+$00
is at  a distance of 40 pc, while the warmer LAS sample lies 
140--200 pc distant. Because of the large distances, the implied tangential 
velocities for the LAS sample are also much higher than that of the SDSS 
white dwarf: 70--120 km~s$^{-1}$ cf.\ 30 km~s$^{-1}$.
Allowing for a range in gravity, the LAS white dwarfs may be $\sim$30 pc closer or
more distant, which translates into a range of velocities of 60--140 km~s$^{-1}$
(Table 4). The sense of the gravity effect is that more massive white dwarfs 
will have a longer cooling age and be closer and slower, and vice versa. 
Even allowing for a generous range in mass for our LAS white dwarf sample,
their distances and motions remain large. Parallax determination will be difficult
for these white dwarfs.

The galaxy simulations of Robin et al.\ (2003) and 
Haywood, Robin \& Cr\'{e}z\'{e} (1997) predict scale heights, velocity 
dispersions and ages for the thin and thick disk and stellar halo 
(or spheroid) components of our Galaxy. The ages of these three components
are 0 -- 10 Gyr, 11 Gyr and 14 Gyr, respectively; the $UVW$ dispersions are 
20 -- 40 km~s$^{-1}$ for thin disk stars older than 3~Gyr, 40 -- 70 km~s$^{-1}$ 
for the thick disk, and 80 -- 130 km~s$^{-1}$ for the halo. The scale heights 
are 100 -- 160 pc for the thin disk, and 800 pc for the thick disk.  

The high velocities of the $T_{\rm eff} \sim$ 6000~K LAS white dwarfs are 
suggestive of thick disk or even halo membership.  Given the short cooling age,
this implies that they would then be remnants of thick disk or halo late-F or 
G stars.  Alternatively, they may be younger thin disk remnants with high 
velocities, such as described in Bergeron (2003).
Conversely, SDSS J2242$+$00 appears to be old and nearby, 
with a low space motion. The velocities could be consistent with
thick disk membership if there is a significant radial component. A parallax for 
this white dwarf would be helpful for further analysis.

Although our $J - H$ selection was designed to pick up $T_{\rm eff} < 4000$~K sources 
(Figure 2), only one was found, together with significantly warmer sources. 
Pushing the LAS to its limits led to large uncertainties in the H magnitude,
hence warmer white dwarfs were scattered into our catalog selection.
The volume probed by Kilic et al.\ (2006) in their search of 3320~deg$^2$ of
the second Data Release of the SDSS, to $g \approx $20 mag, is approximately
three times larger than the volume probed here (the LAS DR2 area is around a 
factor of 12 smaller, while we reached a depth $\sim 1.6\times$ larger). 
Kilic et al.\ found seven white dwarfs with $T_{\rm eff} < 4000$~K, hence we might have 
expected to find two cool white dwarfs, as opposed to the one found. 
Our color selections could have excluded some very cool
white dwarfs --- see for example the location of 
the possible halo white dwarf (Hall et al.\ 2008) in Figure 2, which is 
redder than both our $g - r$ and $J - H$ upper limits.
Hall et al. state that this white dwarf is redder than predicted by any current model,
indicating the complexity of the physics of these atmospheres.
Having performed this initial search and demonstrated the validity of the technique,
we will now refine our color selections and apply them to more recent and larger
LAS data releases. We expect to discover more of these 
elusive remnants of the early history of the Galaxy.

%
%%%%%%%%%%%%%%%%%%%%%%%%%%%
%%%% Conclusions %%%%
%%%%%%%%%%%%%%%%%%%%%%%%%%%
%
\section{Conclusions}

We have searched 280~deg$^2$ of the second data release of the UKIDSS 
Large Area Survey for cool white dwarfs. Candidates were 
identified by pairing the database with the fifth data release of the Sloan
Digital Sky Survey, and searching for high proper motion stars with neutral 
optical colors and blue near-infrared colors. A 100\% success rate was found 
when we obtained optical spectroscopy of seven candidates; we also recovered 
a previously known cool white dwarf found in the SDSS database; we suggest 
that the remaining two stars in the sample are also white dwarfs.

The newly identified white dwarfs are relatively warm with 
$T_{\rm eff} \approx$ 6000~K\@. Of the seven with spectroscopy, six have 
hydrogen-rich atmospheres and the seventh has a helium-rich atmosphere. Their cooling age
is around 2.5 Gyr. The previously known SDSS white dwarf is cool, with a 
mixed composition atmosphere and $T_{\rm eff}$ = 3800~K; the cooling age 
is correspondingly larger at 9 Gyr. Our data does not allow us to constrain
surface gravity or mass for our sample, and we cannot determine 
total age. The cooling age and estimated tangential velocity of the
coolest object suggests that it is an old member of the  disk of the Galaxy,
with an age 10 -- 11~Gyr. The warmer white dwarfs have smaller cooling ages and
higher estimated velocities -- they may be remnants of low-mass stars and 
therefore 11~Gyr-old members of the thick disk,  or they may  be $\sim$5~Gyr-old
thin disk remnants with high velocities. 

We will expand this sample with continued larger-area data releases of the 
UKIDSS LAS. Based on the results presented here, we will refine our color 
selection and in the next data release we should find several 
cool -- and therefore necessarily old -- white dwarf remnants of early star 
formation in the thick disk or even of the halo of the Galaxy.

%
%%%%%%%%%%%%%%%%%%%%%%%%%%%
%%%% Thanks! %%%%
%%%%%%%%%%%%%%%%%%%%%%%%%%%
%
\acknowledgments

We are grateful to the referee for comments which significantly improved the paper.
We thank M.\ Kilic for sending us the electronic form of
his published spectrum. SKL and AN are supported by the Gemini Observatory,
which is operated by the Association of Universities for Research in
Astronomy, Inc., on behalf of the international Gemini partnership. PB
is a Cottrell Scholar of Research Corporation and he is supported by
the NSERC Canada and by the Fund FQRNT (Qu\'ebec). This research has
made use of the Simbad database of NASA's Astrophysics Data System
Bibliographic Services (ADS).  The United Kingdom Infrared Telescope
is operated by the Joint Astronomy Centre on behalf of the
Science and Technology Facilities Council of the U.K. Based on
observations obtained at the Gemini Observatory (program
GN-2007B-DD-6), which is operated by the Association of Universities
for Research in Astronomy, Inc., under a cooperative agreement with
the NSF on behalf of the Gemini partnership: the National Science
Foundation (United States), 
the Science and Technology Facilities Council (United Kingdom), the
National Research Council (Canada), CONICYT (Chile), the Australian Research Council
(Australia), Ministério da Ciência e Tecnologia (Brazil) and SECYT (Argentina).
The SDSS is
managed by the Astrophysical Research Consortium for the Participating
Institutions. The Participating Institutions are the American Museum
of Natural History, Astrophysical Institute Potsdam, University of
Basel, University of Cambridge, Case Western Reserve University,
University of Chicago, Drexel University, Fermilab, the Institute for
Advanced Study, the Japan Participation Group, Johns Hopkins
University, the Joint Institute for Nuclear Astrophysics, the Kavli
Institute for Particle Astrophysics and Cosmology, the Korean
Scientist Group, the Chinese Academy of Sciences (LAMOST), Los Alamos
National Laboratory, the Max-Planck-Institute for Astronomy (MPIA),
the Max-Planck-Institute for Astrophysics (MPA), New Mexico State
University, Ohio State University, University of Pittsburgh,
University of Portsmouth, Princeton University, the United States
Naval Observatory, and the University of Washington.


%% in parentheses, after the keyword, but they will not be verified.

{\it Facilities: \facility{UKIRT (WFCAM, UKIDSS LAS DR2 \& DR3)} \facility{Sloan (SDSS DR5 \& DR6)}, \facility{Gemini:Gillett (GMOS-N)}}



%
%%%%%%%%%%%%%%%%%%%%%%%%%%%
%%%% Bibliography %%%%
%%%%%%%%%%%%%%%%%%%%%%%%%%%
%

%
%%%%%%%%%%%%%%%%%%%%%%%%%%%
%%%% Table: Objects %%%%
%%%%%%%%%%%%%%%%%%%%%%%%%%%
%
\begin{deluxetable}{lcccccc}
%\tabletypesize{\scriptsize}
\tabletypesize{\footnotesize}
%\tablecolumns{7}
\tablewidth{0pt}
%\rotate
\tablecaption{Astrometry for Candidate White Dwarfs in DR2 of the UKIDSS LAS}
\tablehead{
\colhead{Short Name} & \colhead{Right Ascension} & \colhead{Declination} & \colhead{Epoch} & 
\colhead{$\mu$ $\arcsec$yr$^{-1}$} & \colhead{RPM}\\
\colhead{} & \colhead{HH:MM:SS.SS} & \colhead{DD:MM:SS.S} & \colhead{YYYYMMDD} & \colhead{} & \colhead{} & \colhead{}\\
\colhead{} & \colhead{} & \colhead{} & \colhead{}
& \colhead{(RA,dec)} & \colhead{$H_g$} \\
}
\startdata
ULAS J0049$-$00{\tablenotemark{a}} & 00:49:00.53 & $-$00:39:42.1 & 20050902  & $-$0.120,$-$0.034 & 20.36 \\
ULAS J0142$+$00{\tablenotemark{a}} & 01:42:21.79 & $+$00:35:50.9 & 20051126  & $+$0.050,$+$0.092 & 20.36 \\
ULAS J0226$-$00{\tablenotemark{a}} & 02:26:26.53 & $-$00:39:34.9 & 20050926  & $-$0.049,$-$0.098 & 20.85 \\
ULAS J0302$+$00{\tablenotemark{a}} & 03:02:21.35 & $+$00:55:57.0 & 20051007  & $+$0.118,$-$0.031 & 21.33 \\
ULAS J1522$+$08{\tablenotemark{a}} & 15:22:29.87 & $+$08:12:13.9 & 20060703  & $+$0.033,$-$0.099 & 21.04 \\
ULAS J1528$+$06{\tablenotemark{b}} & 15:28:07.15 & $+$06:04:59.4 & 20050528  & $-$0.061,$+$0.083 & 20.94 \\
ULAS J1554$+$08{\tablenotemark{b}} & 15:54:31.37 & $+$08:02:48.5 & 20060723  & $-$0.081,$-$0.113 & 20.91 \\
SDSS J2242$+$00{\tablenotemark{c}} & 22:42:06.23 & $+$00:48:22.4 & 20051007  & $+$0.142,$-$0.084 & 20.76 \\
ULAS J2331$-$00{\tablenotemark{a}} & 23:31:47.60 & $-$00:48:50.0 & 20050828  & $+$0.137,$+$0.003 & 21.16 \\
ULAS J2339$-$00{\tablenotemark{a}} & 23:39:41.65 & $-$00:43:06.4 & 20050828  & $+$0.099,$+$0.029 & 20.49 \\
\enddata
\tablecomments{Typical uncertainty in proper motion is 14 mas yr$^{-1}$.}
\tablenotetext{a}{Confirmed as a white dwarf spectroscopically in this work.}
\tablenotetext{b}{Unconfirmed as a white dwarf.}
\tablenotetext{c}{Discovered in SDSS by Kilic et al. (2006) and confirmed 
spectroscopically by those authors. Our proper motion determination is in 
agreement with their estimate.}
\end{deluxetable}

\clearpage

\clearpage

%
%%%%%%%%%%%%%%%%%%%%%%%%%%%
%%%% Table: Photometry %%%%
%%%%%%%%%%%%%%%%%%%%%%%%%%%
%
\begin{deluxetable}{lrrrrrrrr}
\tabletypesize{\scriptsize}
%\tablecolumns{9}
\tablewidth{0pt}
\rotate
\tablecaption{Photometry for Candidate White Dwarfs in DR2 of the UKIDSS LAS}
\tablehead{
\colhead{Short Name} & \colhead{$u$(err)} & \colhead{$g$(err)} & \colhead{$r$(err)} & \colhead{$i$(err)} &
\colhead{$z$(err)} & \colhead{$Y$(err)}  & \colhead{$J$(err)} & \colhead{$H$(err)}   \\
}
\startdata
ULAS J0049$-$00{\tablenotemark{a}} & 20.78(0.11) &  19.87(0.02) &  19.52(0.02) &  19.42(0.02) &  19.50(0.09) 
& 19.05(0.08)  & 18.72(0.11) &  18.82(0.27) \\
ULAS J0142$+$00{\tablenotemark{a}} & 20.96(0.09) &  20.26(0.02) &  19.86(0.02) &  19.68(0.03) &  19.68(0.08)
& 19.21(0.08)  & 18.75(0.09) &  18.94(0.21) \\
ULAS J0226$-$00{\tablenotemark{a}} & 21.83(0.22) &  20.66(0.03) &  20.14(0.02) &  20.03(0.03) &  19.97(0.09)
& 19.36(0.10)  & 18.98(0.14) &  19.08(0.23) \\
ULAS J0302$+$00{\tablenotemark{a}} & 21.66(0.18) &  20.90(0.03) &  20.48(0.03) &  20.27(0.04) &  20.23(0.13)
& 19.60(0.11)  & 19.28(0.16) &  19.42(0.30)  \\
ULAS J1522$+$08{\tablenotemark{a}} & 22.44(0.39) &  20.95(0.04) &  20.48(0.03) &  20.19(0.04) &  19.86(0.08)
& 19.55(0.11)  & 19.19(0.12) &  19.31(0.24)  \\
ULAS J1528$+$06{\tablenotemark{b}} & 21.96(0.22) &  20.88(0.03) &  20.57(0.03) &  20.37(0.04) &  20.48(0.14)
& 19.78(0.10) &  19.52(0.15) & 19.71(0.31)  \\
ULAS J1554$+$08{\tablenotemark{b}} & 21.15(0.09) &  20.20(0.02) &  19.88(0.02) &  19.77(0.02) &  19.86(0.08)
&  19.18(0.07) & 18.75(0.09) &  18.96(0.17)  \\
SDSS J2242$+$00{\tablenotemark{c}} & 22.26(0.25) &  19.66(0.01) &  18.65(0.01) &  18.28(0.01) &  18.16(0.02)
& 17.71(0.02) & 18.02(0.05) & 18.59(0.15) \\
ULAS J2331$-$00{\tablenotemark{a}} & 21.47(0.19) &  20.48(0.03) &  20.15(0.03) &  20.06(0.04) &  19.87(0.12)
& 19.55(0.11)  & 19.23(0.15) &  19.43(0.29) \\
ULAS J2339$-$00{\tablenotemark{a}} &  21.33(0.18) & 20.40(0.03) &  20.17(0.03) &  20.05(0.04) &  19.94(0.12)
& 19.48(0.09)  & 19.24(0.13) &  19.39(0.27)  \\
\enddata
\tablecomments{None of the sources were detected at $K$, implying $K >$ 18.2
mag. SDSS DR6 (Adelman-McCarthy et al.\ 2008) and LAS DR3 photometry is given
although our candidates were selected from LAS DR2 and SDSS DR5. SDSS 
$ugriz$ magnitudes are on the AB system (Fukugita et al. \ 1996)
while LAS $YJHK$ are on the Vega
system (Hewett et al.\ 2006).}
\tablenotetext{a}{Confirmed as a white dwarf spectroscopically in this work.}
\tablenotetext{b}{Unconfirmed as a white dwarf.}
\tablenotetext{c}{Discovered in SDSS by Kilic et al.\ (2006) and 
confirmed spectroscopically by those authors.}
\end{deluxetable}

\clearpage

%
%%%%%%%%%%%%%%%%%%%%%%%%%%%%%%%%%%%
%%%% Table: Spectroscopic Logs %%%%
%%%%%%%%%%%%%%%%%%%%%%%%%%%%%%%%%%%
%
\begin{deluxetable}{lccc}
\tablecaption{GMOS-N Observation Log}
\tablewidth{0pt}
\tablehead{
\colhead{Short Name} & \colhead{SDSS $r$} & \colhead{Total Exp} & \colhead{Dates} \\
\colhead{}  & \colhead{AB}  &\colhead{seconds} & \colhead{YYYYMMDD} \\
}
\startdata
ULAS J0049$-$00 & 19.56 & 600  & 20080109\\
ULAS J0142$+$00 & 19.86 & 2400 & 20080109\\
ULAS J0226$-$00 & 20.14 & 4200 & 20080109\\
ULAS J0302$+$00 & 20.48 & 6000 & 20080103, 20080110\\
ULAS J1522$+$08 & 20.48 & 6000 & 20080304\\
ULAS J2331$-$00 & 20.15 & 4200 & 20080103\\
ULAS J2339$-$00 & 20.17 & 4200 & 20080110\\
\enddata
\label{tab:ObsLog}
\end{deluxetable}

\clearpage

%
%%%%%%%%%%%%%%%%%%%%%%%%%%%%%
%%%% Table: Properties %%%%
%%%%%%%%%%%%%%%%%%%%%%%%%%%%%
%
\begin{deluxetable}{lcccrr}
\tabletypesize{\footnotesize}
%\tablecolumns{6}
\tablewidth{0pt}
\tablecaption{Derived Properties for the DR2 UKIDSS/LAS White Dwarfs}
\tablehead{
\colhead{Short Name} & \colhead{Atmospheric{\tablenotemark{a}}} & \colhead{$T_{\rm eff}$}
& \colhead{Cooling{\tablenotemark{b}}} &  \colhead{Distance{\tablenotemark{c}}}
& \colhead{$V_{\rm tan}${\tablenotemark{d}}}\\
\colhead{}  & \colhead{Composition}  &\colhead{K}
& \colhead{Age, Gyr} & \colhead{pc} & \colhead{km~s$^{-1}$}\\
}
\startdata
ULAS J0049$-$00{\tablenotemark{e}} & H & 6380$\pm$140 & 1.9$\,^{+\,1.8}_{-\,0.6}$ & 140$\pm$25 & 80$\pm$15 \\
ULAS J0142$+$00{\tablenotemark{e}} & H & 5950$\pm$140 & 2.3$\,^{+\,2.1}_{-\,0.8}$ & 140$\pm$25 & 70$\pm$15 \\
ULAS J0226$-$00{\tablenotemark{e}} & H & 5670$\pm$160 & 2.7$\,^{+\,2.5}_{-\,1.0}$ & 150$\pm$30 & 80$\pm$15 \\
ULAS J0302$+$00{\tablenotemark{e}} & He & 5720$\pm$160 & 3.2$\,^{+\,2.2}_{-\,1.3}$ & 180$\pm$35 & 100$\pm$20 \\
ULAS J1522$+$08{\tablenotemark{e}} & H & 5390$\pm$180 & 3.6$\,^{+\,3.0}_{-\,1.6}$ & 150$\pm$30 & 70$\pm$15 \\
ULAS J1528$+$06{\tablenotemark{f}} & unknown & (6060 - 6160)$\pm$200 &  (2.1 - 2.5)$\,^{+\,2.1}_{-\,0.7}$ & 200$\pm$40 & 100$\pm$20 \\
ULAS J1554$+$08{\tablenotemark{f}} & unknown & (6210 - 6330)$\pm$140 &  (2.0 - 2.3)$\,^{+\,1.9}_{-\,0.7}$ & 160$\pm$30 & 100$\pm$20 \\
SDSS J2242$+$00{\tablenotemark{g}} & H$\simeq$He  & 3820$\pm$100 & 8.7$\,^{+\,0.4}_{-\,1.6}$ &  37$\pm$6 & 29$\pm$5 \\
ULAS J2331$-$00{\tablenotemark{e}} & H & 6340$\pm$220 & 1.9$\,^{+\,1.8}_{-\,0.6}$ & 180$\pm$20 & 120$\pm$20 \\
ULAS J2339$-$00{\tablenotemark{e}} & H & 6590$\pm$240 & 1.8$\,^{+\,1.5}_{-\,0.6}$ & 190$\pm$35 & 90$\pm$20 \\
\enddata
\tablecomments{A standard surface gravity of $\log g=8.0$ has been adopted; 
this does not significantly affect the derived composition or temperature,
but does impact age, distance and velocity. For these last three parameters a range is shown corresponding to 
$7.7 \leq \log g \leq 8.3$.}
\tablenotetext{a}{The atmospheric composition is by number.}
\tablenotetext{b}{The cooling age is derived from the composition and temperature using the cooling models of Fontaine, Brassard \& Bergeron (2001). }
\tablenotetext{c}{The distance is estimated from the modelled and observed 
magnitudes (Holberg \& Bergeron 2006). }
\tablenotetext{d}{The tangential velocity is calculated from distance and
proper motion given in Table 1.}
\tablenotetext{e}{Confirmed as a white dwarf spectroscopically in this work.}
\tablenotetext{f}{Unconfirmed as a white dwarf.}
\tablenotetext{g}{Discovered in SDSS by Kilic et al.\ (2006). }
\end{deluxetable}

\clearpage

%
%%%%%%%%%%%%%%%%%%%%%%%%%%%%%
%%%%  astrometry plot 
%%%%
%%%%%%%%%%%%%%%%%%%%%%%%%%%%%
%
\begin{figure}
\begin{center}
\includegraphics[angle=0,scale=.60]{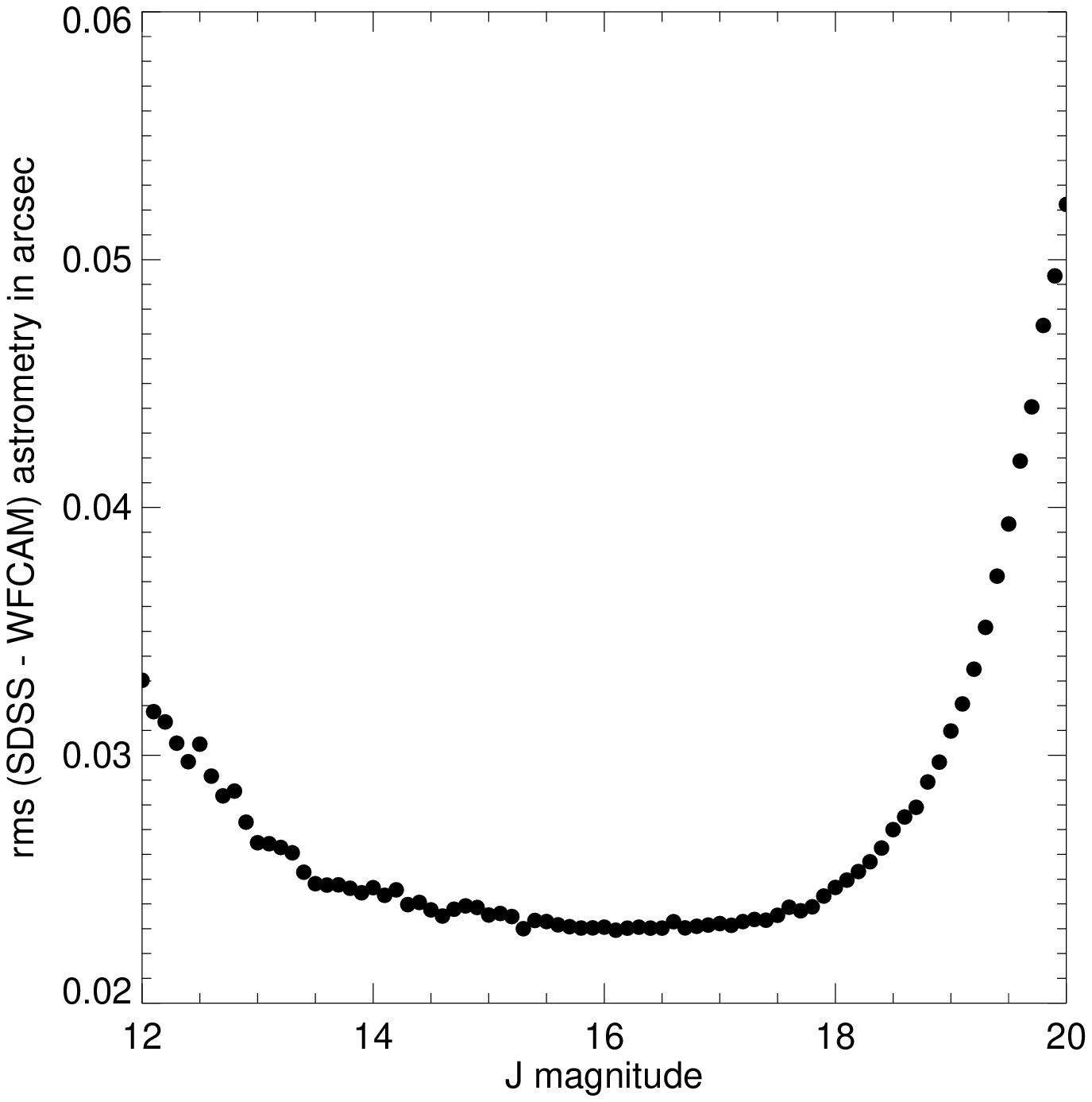}
\end{center}
\caption{The $rms$ of the difference in astrometry for all point sources 
in LAS DR2 and SDSS DR5 as a function of $J$ magnitude.
}
\end{figure}

%
%%%%%%%%%%%%%%%%%%%%%%%%%%%%%
%%%% Figure: CC-Plot %%%%
%%%%%%%%%%%%%%%%%%%%%%%%%%%%%
%
\begin{figure}
\begin{center}
\includegraphics[angle=0,scale=.53]{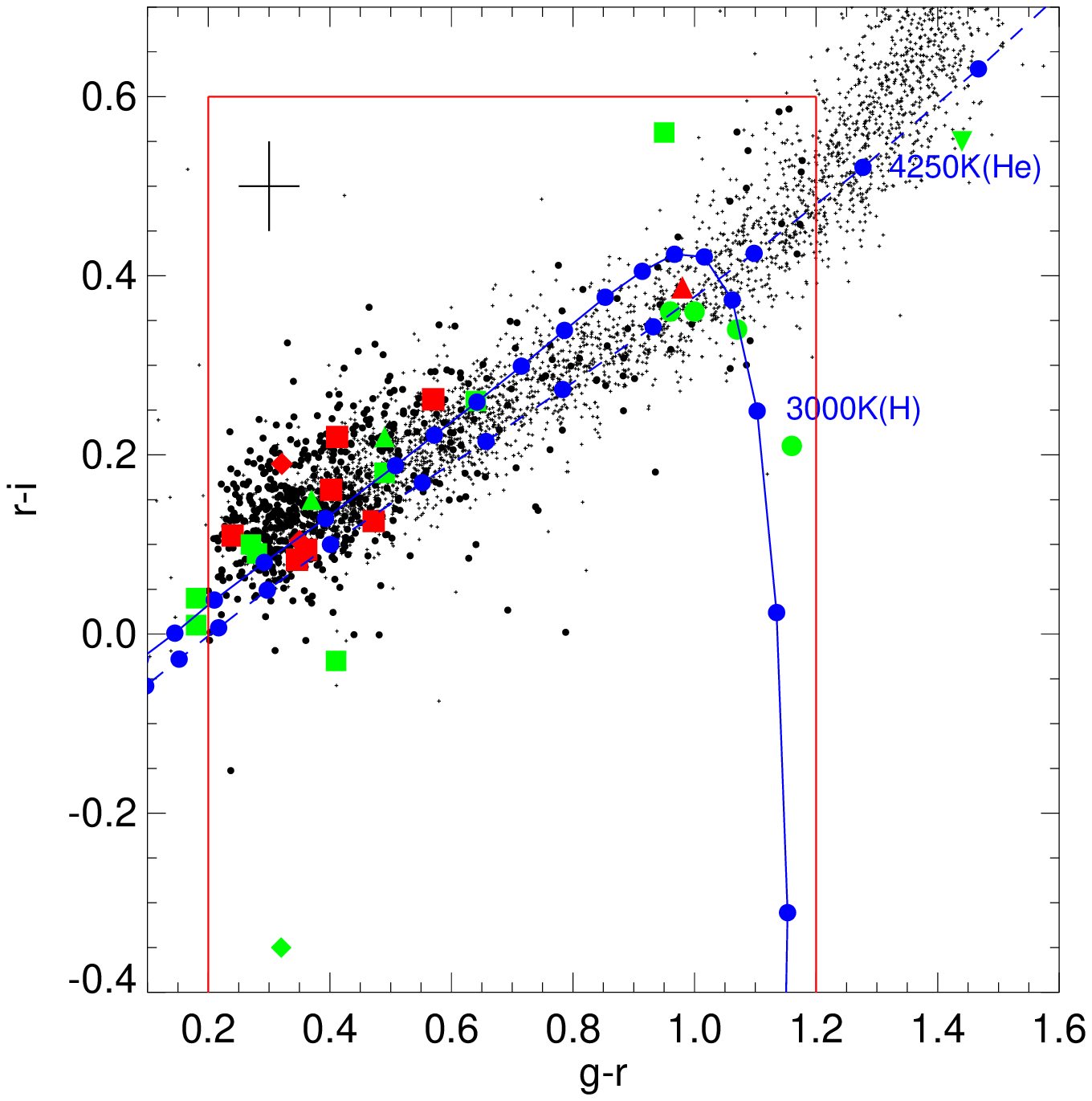}
\includegraphics[angle=0,scale=.53]{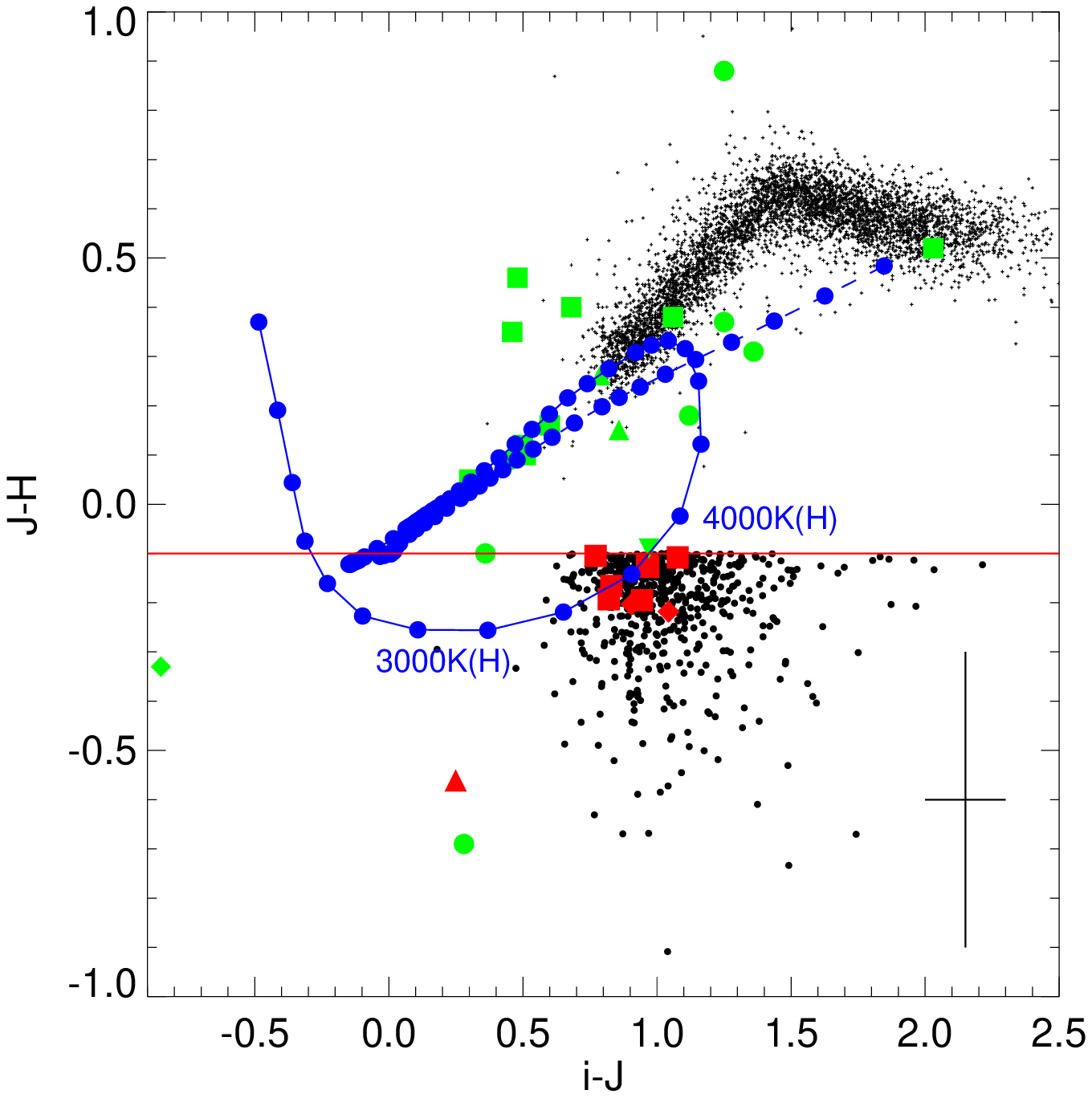}
\end{center}
\caption{Color-color plots for our  seven new confirmed white dwarfs (red squares), 
one recovered Kilic et al. \ (2006) white dwarf (red triangle),
two candidate white dwarfs (red diamonds), and other white dwarfs 
taken from the literature (green symbols; triangles - Kilic et al. \ 2006;
diamond - Harris et al. \ 2001; squares - Carollo et al. \ 2006;
circles - Vidrih et al. \ 2008; downward triangle - Hall et al. \ 2008).   
Model sequences  with $\log g=8$  
(see \S 5.1) are shown for hydrogen (solid lines) and helium (dashed lines) 
atmospheres; $T_{\rm eff}$ decreases from left to right, looping back to 
the left for the hydrogen sequence in the $i - J$:$J - H$ plot. 
Blue dots indicate $\Delta T_{\rm eff} = 250$~K.  Color cuts used to select our
sample are indicated by the red lines. Typical error bars are shown. Also shown 
is the location of the main sequence (small black dots) and the sample selected
on color alone, before the proper motion cut was applied (larger black dots). 
$gri$ are on the AB system, $JH$ on the Vega system.}
\end{figure}

\clearpage

%
%%%%%%%%%%%%%%%%%%%%%%%%%%%%%
%%%% Figure: RPM %%%%
%%%%%%%%%%%%%%%%%%%%%%%%%%%%%
%
\begin{figure}
\includegraphics[angle=0,scale=.60]{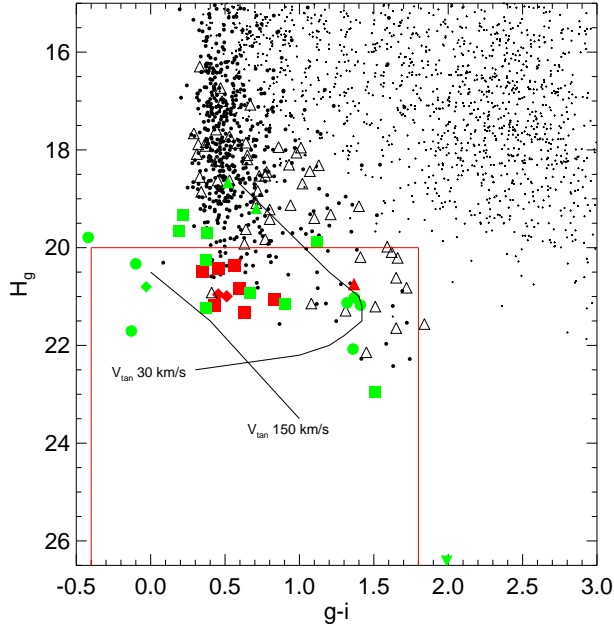}
\caption{The reduced $g$-band proper motion as a function of $g - i$ for 
our sample of white dwarfs (red symbols),
and others taken from the literature (green symbols); symbols are as in Figure 2. 
Confirmed subdwarfs from Kilic et al.\ (2006) are 
shown as open  triangles.  The main sequence and our initial candidate selection
based on color alone are also shown, as small and large black dots respectively.
Red lines indicate the region included by our color and proper motion cuts.
White dwarf cooling curves for different tangential velocities are shown 
as solid lines. The 30 km~s$^{-1}$ curve marks the expected location of disk 
white dwarfs, and the 150 km~s$^{-1}$ curve represents the halo white dwarfs.
}
\label{fig:RPM}
\end{figure}

\clearpage

%
%%%%%%%%%%%%%%%%%%%%%%%%%%%%%
%%%% Figure: Spectra %%%%
%%%%%%%%%%%%%%%%%%%%%%%%%%%%%
%
\begin{figure}
\includegraphics[angle=-90,scale=.60]{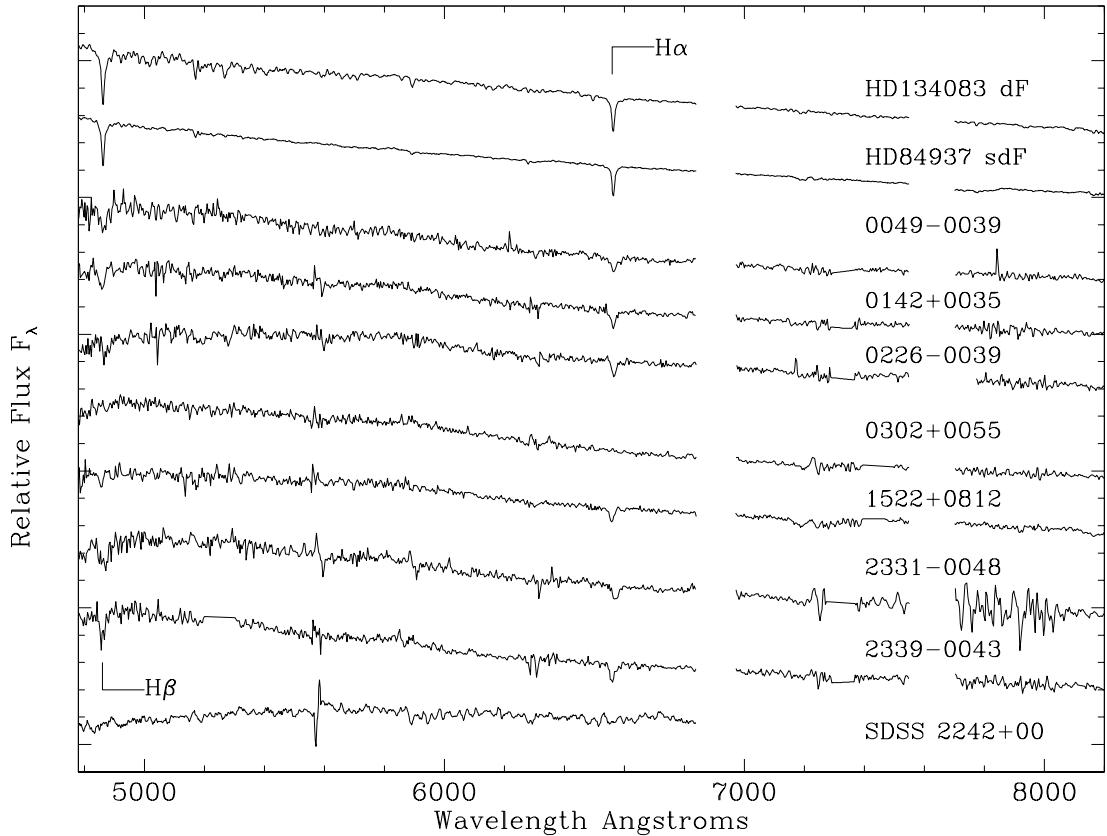}
\caption{The spectra obtained with GMOS-N at Gemini for the sample of seven 
white dwarfs. Exposure times are given in Table \ref{tab:ObsLog}, and the 
instrument configuration is described in the text.  The telluric oxygen 
A and B lines at 6867 and 7594\AA{} have been removed for clarity. The 
linear regions seen for some white dwarfs are the gaps in the GMOS-N detectors. 
F dwarf and F subdwarf spectra 
(Le Borgne et al.\ 2003) are shown at the top of the panel for comparison, 
where the spectra have been smoothed to the resolution of our data. The spectrum 
at the bottom of the panel is  from Kilic et al.\ (2006) and is the cool SDSS
white dwarf discovered by those authors and recovered in this work.
}
\label{fig:GMOSspectra}
\end{figure}

\clearpage

%
%%%%%%%%%%%%%%%%%%%%%%%%%%%%%
%%%% Figure: Fits #1 %%%%
%%%%%%%%%%%%%%%%%%%%%%%%%%%%%
%
\begin{figure}
\begin{center}
\includegraphics[angle=0,scale=.65]{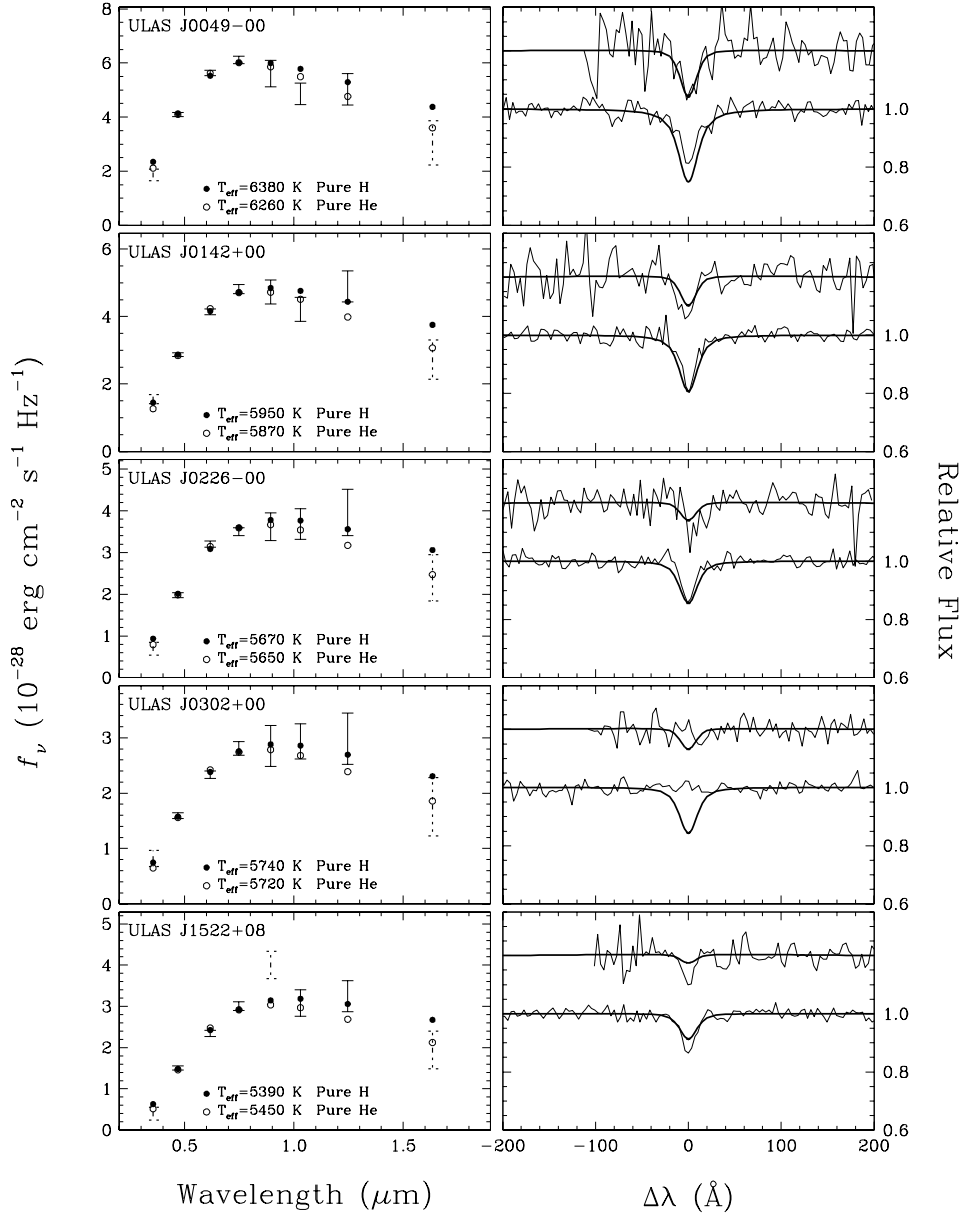}
\end{center}
\caption{Plots demonstrating the model fits to five of the white dwarfs in
our sample, as identified in the legends. The error bars in the left panels
represent  SDSS and LAS photometry; SDSS $u$ (and $z$ for ULAS J1522$+$08),
 and  LAS $H$,
have been ignored in the fits (dashed error bars). 
Circles represent the models fluxes averaged over the filter bandpass;
filled circles are pure-hydrogen models and open circles are pure-helium models.
A surface gravity $\log g=8.0$ is
assumed, and the derived $T_{\rm eff}$ for each composition is shown.
The right panels show the observed spectrum around the H$\beta$ (top) and 
H$\alpha$ (bottom) lines, with the modelled pure-hydrogen atmosphere line 
profiles. ULAS J0302$+$00 has a featureless spectrum, and is therefore helium-rich;
the remaining sources are hydrogen-rich.
} 
\label{fig:WDfits_1}
\end{figure}

\clearpage

%
%%%%%%%%%%%%%%%%%%%%%%%%%%%%%
%%%% Figure: Fits #2 %%%%
%%%%%%%%%%%%%%%%%%%%%%%%%%%%%
%
\begin{figure}
\begin{center}
\includegraphics[angle=0,scale=.65]{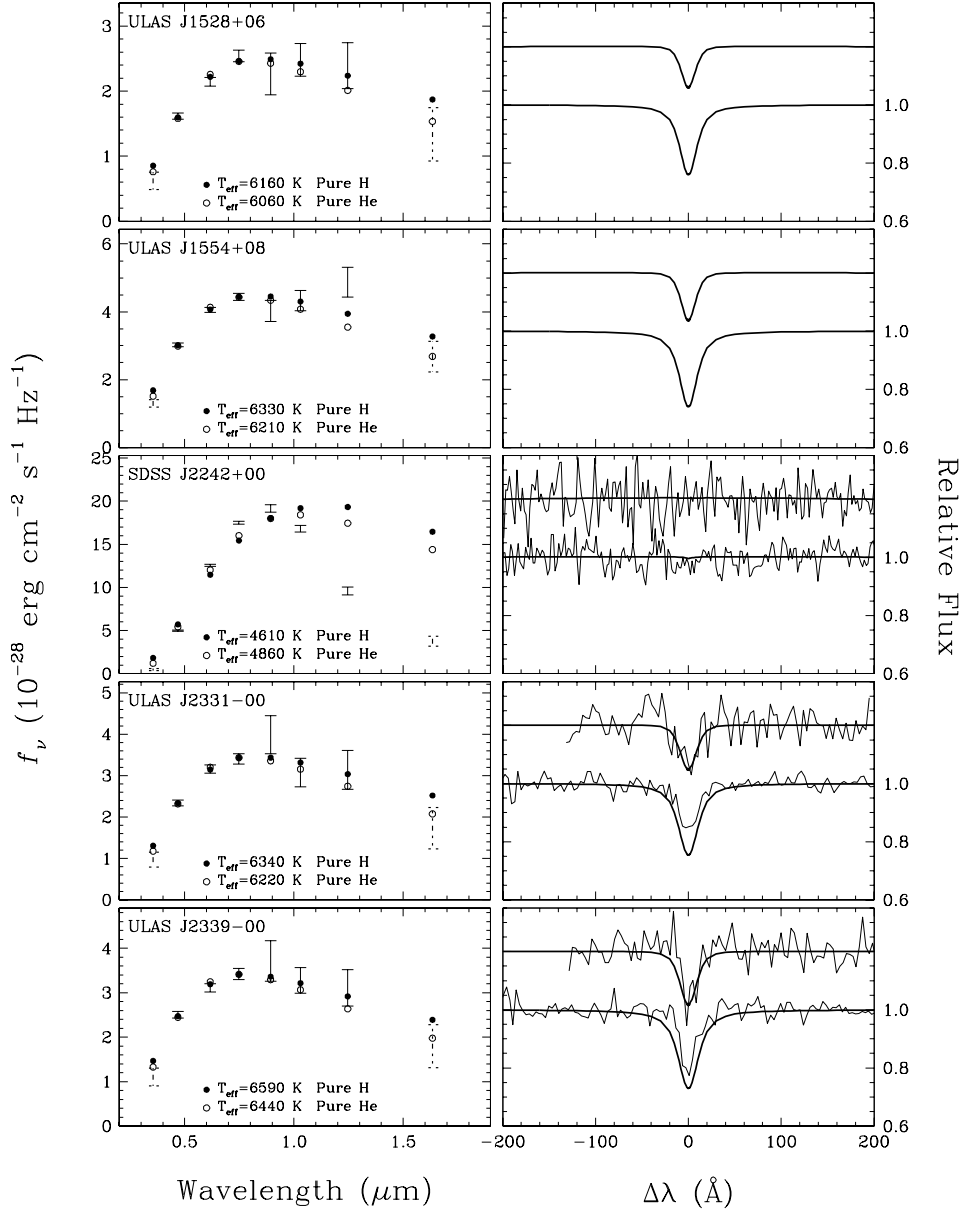}
\end{center}
\caption{Similar to Figure \ref{fig:WDfits_1}, for the remaining white dwarfs 
in our sample. For ULAS J1528$+$06 and ULAS J1554$+$08 no spectra exists and 
only modelled H$\beta$ and H$\alpha$ spectra are shown. For SDSS J2242$+$00 
both the pure-hydrogen and pure-helium fits to the photometry are poor; 
the modelled pure-hydrogen spectra shown are featureless as no absorption 
lines would be detected at 4610~K (a better fit to this object is shown in 
Figure \ref{fig:fit_SDSS2242}).
}
\label{fig:WDfits_2}
\end{figure}

\clearpage

%
%%%%%%%%%%%%%%%%%%%%%%%%%%%%%
%%%% Figure: Fits #3 %%%%
%%%%%%%%%%%%%%%%%%%%%%%%%%%%%
%
\begin{figure}
\includegraphics[angle=0,scale=.70]{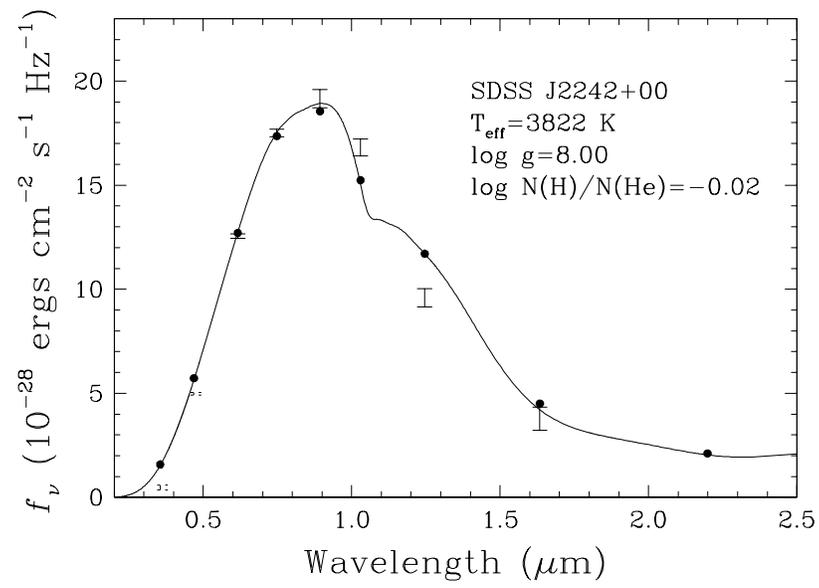}
\caption{Model fits to SDSS J2242$+$00; this mixed-composition atmosphere 
fit is superior to the single composition fits shown in 
Figure \ref{fig:WDfits_2}\@.
}
\label{fig:fit_SDSS2242}
\end{figure}
\clearpage

\end{document}